# Economic and Technical Feasibility of V2G in Non-Road Mobile Machinery sector

**Nicolas Rößler[1], Irfan Khan[2], Thomas Schade[2], Christoph Wellmann[1], Xinyuan Cao[1], Milan Kopynske[1], Feihong Xia[2], Rene Savelsberg[2], Univ.-Prof. Jakob Andert[1]**

*[1]RWTH Aachen University, MMP - Teaching and research area mechatronics in mobile propulsion*
*Forckenbeckstraße 4, 52074 Aachen, Germany (E-mail: roessler@mmp.rwth-aachen.de)*

*[2]FEV Europe GmbH*
*Neuenhofstraße 181, 52078 Aachen, Germany (E-mail: khan_i@fev.com)*

**ABSTRACT:** This paper investigates the economic and technical feasibility of integrating Vehicle-to-Grid (V2G) technology in the Non-Road Mobile Machinery (NRMM) sector. These often-idling assets, with their substantial battery capacities, present a unique opportunity to participate in energy markets, providing grid services and generating additional revenue. A novel methodology is introduced that integrates Bayesian Optimization (BO) to optimize the energy infrastructure together with an operating strategy optimization to reduce the electricity costs while enhancing grid interaction. While the focus lies on the methodology, the financial opportunities for the use-case of an electric NRMM rental service will be presented. However, the study is limited by the availability of real-world data on the usage of electric NRMM and does not address regulatory challenges of V2G. Further research is needed to extend the model accuracy and validate these findings.

## 1. INTRODUCTION

Vehicle-to-Grid (V2G) technology has emerged as a key enabler in the transition to more flexible, resilient, and sustainable energy systems [1]. By allowing a bidirectional energy flow between electric vehicles (EVs) and the power grid, V2G offers the potential to support grid stability, mitigate peak demand, and provide valuable ancillary services [2]. While V2G is frequently emphasized in discussions about grid services, the concept also enables vehicle to home (V2H), building (V2B) and load (V2L) [3] , and can additionally serve as a backup power source during grid outages [4]. This versatility supports decentralized energy systems and offers practical benefits for various user groups, including rural and industrial applications.

Building upon existing V2G applications in passenger vehicles and fleets, this paper turns attention to a novel use case: the integration of V2G with electric Non-Road Mobile Machinery (NRMM). Therefore, the application of V2G within a schedulable rental system for electric NRMM is explored. The aim is to explore how extending the application of V2X to electric NRMM can support energy cost optimization through price-aware charging strategies and help reduce peak loads on the grid. By extending the scope of V2G beyond traditional EVs, this approach highlights new opportunities for smart energy management in industrial and construction settings.

In agricultural farms, where energy demand is particularly high due to the use of mobile machinery, heating, and irrigation systems, the integration of electric NRMM with bidirectional charging functionality (V2B, V2L, V2G) could play a key role in improving energy flexibility and resilience [5]. Reflecting this potential, the electrification of machinery such as tractors and construction vehicles is gaining traction, driven by the need for reduced carbon emissions and technological advancement. Studies show that transitioning to electric NRMM could significantly lower their environmental impact, providing both operational and sustainability benefits. Moreover, electric NRMM align with broader energy transition goals by reducing reliance on fossil fuels and enabling cleaner energy use. These advancements not only help to address pressing environmental concerns but also contribute to the modernization of industrial fleets [6, 7].

This shift is increasingly accompanied by digital transformation processes: IoT-based monitoring and energy management systems are being adopted in agriculture, providing the infrastructural foundation for integrating V2G into automated operational workflows [8]. Initial pilot projects in the agricultural and construction sectors have begun to explore the use of electric NRMM in combination with smart charging and grid services, underlining the growing practical relevance of such approaches [9]. As V2G technology gains momentum in various sectors, its application to NRMM fleets holds potential to further enhance these environmental and operational benefits.

While several barriers to V2G implementation in private cars, such as range anxiety and the need for predictable parking periods, have been identified [10], these issues are less pronounced in the case of NRMM. Thanks to their predictability of usage and the high idle time [11], both range anxiety and parking requirements for V2G are less of a concern. Research has already investigated the potential of V2G for fleets of electric busses [12, 13]. These cases share important characteristics with the system proposed in this paper, such as predictable schedules and centralized charging infrastructure, which enable effective energy planning and grid interaction. The flattening of peak loads and the economic optimization of bidirectional charging have proven effective, which suggests that similar mechanisms could also be successfully applied in the context of an NRMM fleet.

A recent study has demonstrated that certain types of electric NRMM, such as tractors, can achieve a lower total cost of ownership (TCO) compared to their diesel counterparts over the long term [14]. Moreover, a study on EV fleets at workplaces, integrated with PV and utilizing V2G charging, found a reduction in net costs by 32%, further emphasizing the economic benefits of integrating V2G into fleet management [15]. These findings indicate that electrified NRMM is not only environmentally beneficial but also economically viable, especially when combined with energy management strategies such as V2G.

Beyond general industrial use, the agricultural sector represents a particularly promising domain for V2G-capable electric NRMM - especially tractors. When combined with on-site photovoltaic systems, these vehicles can store surplus solar energy and reduce electricity costs, increasing energy autonomy on farms. A case study from a rural village in India illustrates how such systems - when owned by farmers - can support sustainable agricultural practices through improved self-sufficiency and micro cold storage integration [16]. While this paper does not focus on privately owned machinery, the underlying energy management potential also applies to rental-based models. In these cases, pooled electric NRMM fleets can offer comparable grid and cost benefits through coordinated V2G strategies, demonstrating the flexibility of the approach across different ownership structures.

## 2. METHODOLOGY

In this study, a simulation of a fleet rental service is conducted aiming to reduce its electricity cost by utilizing an energy management system (EMS) and incorporating bidirectional electric vehicle supply equipment (EVSE). Additionally, the power range of the EVSE, the peak power of a photovoltaic system (PV), the capacity of a battery electric stationary storage (BESS) and the power of the grid connection point will be optimized. A bayesian optimization (BO) will be used to iteratively improve the sizing of these previously mentioned energy components. The simulated scenario with one sizing set will be called inner layer and is described in chapter 2.1 while the optimization of the sizing with the BO will be called outer layer and is described in chapter 2.2. The described system configuration is depicted in Figure 1:

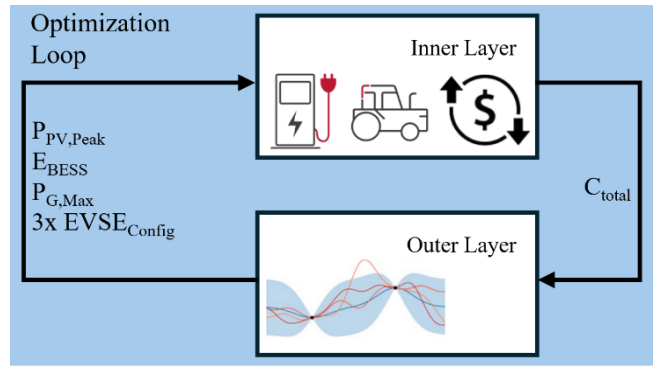

Figure 1: Optimization loop including the inner layer (EMS) and the outer layer (BO)

### 2.1 Inner layer

As mentioned, the model represents a rental company with ten electric NRMM, more specifically tractors, with a corresponding EVSE for each one, and additionally a BESS and a PV system.

The simulation model consists of several sub models, which will be described in the following sections:

- Tractor time & energy demand

- electric models of the BESS, the EVSE, the PV, the dealership house and the grid

- the cost modelling

- the cost optimization with an EMS

#### 2.1.1 Tractor time & energy demand

In order to simulate the total electric consumption of the dealership, the arrival and departure time of the tractors as well as their energy demand are necessary. Therefore, several assumptions are made: All rented tractors need to leave the dealership with a State of Charge (SoC) of 100% and are being returned with a SoC of 5%. Since the main business of the rental service is to rent out tractors, we also assumed a high utilisation rate of the electric vehicles, reducing the time at the dealership. To still have some variety within the ten tractors, three use-cases were determined that repeat weekly, as shown in Figure 2:

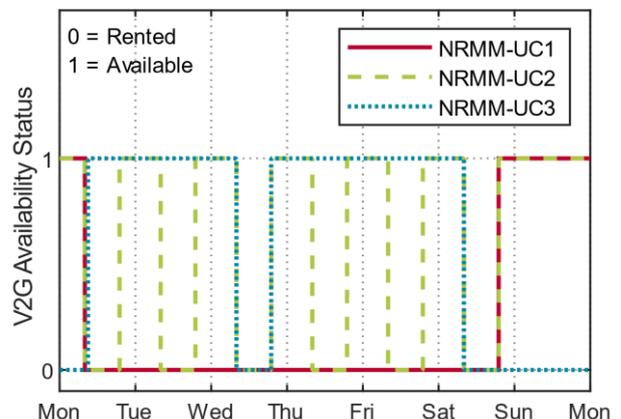

Figure 2: Weekly renting profile of ten NRMMs divided in three use-cases

- Use Case 1: Six vehicles are rented from Monday morning 8am to Saturday evening 7pm.

- Use Case 2: Two vehicles are rented daily (except on Sunday) from 8am to 7pm.

- Use Case 3: Two vehicles are rented on Wednesday and throughout the weekend (from Friday 7pm to Monday 8am).

### 2.1.2 Electric models

The grid connection point and all components drawing or providing power are depicted in Figure 3:

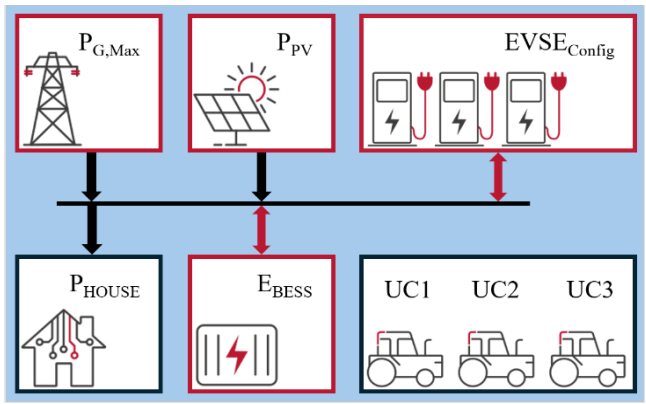

Figure 3: Electric consumers and producers at simulated grid connection point

Based on all the energy components and their power, the power at the grid connection point can be calculated at each step $t$ of the simulation, according to equation (1):

$$\begin{aligned} P_{Grid,p}(t) + P_{Grid,n}(t) &= P_{Grid}(t) \\ &= -P_{PV}(t) + P_{BESS}(t) \\ &\quad + P_{House}(t) + \sum P_{EVSE,i}(t) \end{aligned} \quad (1)$$

One grid connection point and a PV system provides the power for the ten (bidirectional) EVSE charging the tractors, for charging the BESS and for powering the rental service building. The power provided by the photovoltaic system $P_{PV}(t)$ is dependent on the sizing of the PV system and on openly available historic irradiation data from Aldenhoven, Germany – where this rental service shall be located [17]. The orientation and slope of the fixed mounted PV system is optimized for the maximal energy production throughout the year.

Similar to the PV production power, the power of the rental service, called $P_{House}(t)$ is predefined and daily repeating. The mean daily power is set to 1kW, while the peak power is set to 5kW with the highest consumption being in the morning and evening hours. The exact profile of a typical day is later shown in Figure 5.

In addition to the peak PV power, the capacity of the BESS $E_{BESS}$ is also determined by the BO prior to the simulation. The useable SoC range of the BESS is limited to 5%-95%. Additionally, the charging or discharging power of the BESS $P_{BESS}(t)$ is limited with a C-rate of 0.5. These SoC and power limitations ensure that the BESS ageing is minimal. Therefore, capacity or power fades from the BESS throughout the simulation are neglected.

The EVSE configuration is also adaptable by the BO by choosing between eight different charging configurations available for each use-case. Besides the possibility to utilize a bidirectional charging station, also the maximal power for each use-case can be chosen from the following options: 11kW, 22kW, 50kW, 150kW. The charging or discharging power of one EVSE is denoted $P_{EVSE,i}$

The grid power $P_{Grid}(t)$ is divided into the positive $P_{Grid,p}(t)$ and the negative grid power $P_{Grid,n}(t)$. The last unknown parameter set by the BO is the maximal power at the grid connection point: $P_{G,max}$. This limits the maximum $P_{Grid}(t)$ that can be achieved during the simulation, with $abs(P_{Grid}(t)) \leq P_{G,max}$.

### 2.1.3 Cost modeling

The total cost of one setup $c_{Tot}$, used as the objective for the BO are depicted in equations (2), (3) and (4):

$$c_{Tot} = c_{Comp} + \sum c_{Elec}(t) \quad (2)$$

$$\begin{aligned} c_{Comp} = \sum c_{EVSE,i} + c_{PV} + c_{BESS} \\ + c_{Grid,Max} \end{aligned} \quad (3)$$

The total cost, denoted as $c_{Tot}$, is derived from two components: the costs associated with the system components, represented as $c_{Comp}$, which are defined by sizing the energy components and the cumulative costs related to electrical consumption over the course of the simulation, indicated as $c_{Elec}(t)$ .

The component costs include several elements, including the costs for each EVSE $c_{EVSE,i}$, the cost for the PV system $c_{PV}$, the cost for the BESS $c_{BESS}$ and the cost of the maximum grid connection power $c_{Grid,Max}$. Each of these costs are based on the expected lifetime of the component and the assumption that the yearly depreciation is constant. An ageing of the components resulting in peak power reduction or efficiency losses is not considered. More information on the simulated size dependant costs and lifetime of each component can be found in Table 1:

Table 1: Component cost and lifetime

| Components | Costs | Lifetime [years] |
|---|---|---|
| PV | 240 €/kWp | 20 |
| BESS | 200 €/kWh | 10 |
| Grid, max | 106 €/kW | 1 |
| EVSE 11Uni | 3 594 € | 15 |
| EVSE 11Bidi | 4 000 € | 15 |
| EVSE 22Uni | 6 388 € | 15 |
| EVSE 22Bidi | 7 200 € | 15 |
| EVSE 50Uni | 13 500 € | 15 |
| EVSE 50Bidi | 15 345 € | 15 |
| EVSE 150Uni | 38 900 € | 15 |
| EVSE 150Bidi | 44 436 € | 15 |

While the component costs are fixed by the BO at the beginning of the simulation, the electricity costs $c_{Elec}(t)$ depend on the optimization strategy, which is described in chapter 2.1.4.

### 2.1.4 Simulation and optimization

The simulation goal for the inner-layer optimization is the reduction of yearly electricity cost $c_{Elec}$ described in equations (4,5). The simulation is conducted for one year (2023) with MATLAB 24a. In order to reduce the optimization time and due to the weekly repeating NRMM usage, the optimization is performed once for every week. The optimizer is a linear programming solver from MATLABs optimization toolbox with the simplified problem formulation:

$$f = min \sum c_{Elec} \quad (4)$$

$$c_{Elec}(t) = P_{Grid,p}(t) * c_{Buy}(t) * \Delta t \\ + P_{Grid,n}(t) * c_{Feed}(t) * \Delta t \quad (5)$$

The electricity costs $c_{Elec}(t)$ are composed of the cost for buying energy from the grid and the profit from reselling energy to the grid. The bought energy $P_{Grid,p}(t) * \Delta t$ is defined with a positive sign of the grid power $P_{Grid}(t)$, while the sold energy $P_{Grid,neg}(t) * \Delta t$ has a negative sign. The timestep for the simulation is set to $\Delta t = 15 mins$. The dynamic buying tariff $c_{Buy}(t)$ is based on the day-ahead price of the energy market [18]. An exemplary week of this tariff is later depicted in Figure 5. The dynamic feed-in tariff is set to $c_{Feed}(t) = 0.9 * c_{Buy}(t)$. The reasoning behind reducing the feed-back tariff lies within the approach for the optimization and is described later.

By shifting the charging times of the NRMMs and the BESS to a time period with lower grid prices and feeding back power during a time period with higher time periods, the electricity costs can be minimized.

Additionally, several equality and inequality constraints need to be considered. These ensure that equation (1) is valid for each timestep and that the tractors are charged to the desired SoC at departure time. The inequality constraints ensure that the power limits of each component is respected, that the BESS and the vehicles are only charged within their allowed SoC-limits.

While the equality constraint determines the grid power $P_{Grid}(t)$, based on the set BESS and EVSE power; the distribution on the positive power $P_{Grid,p}(t)$ and the negative power $P_{Grid,n}(t)$, shown in equation (1), is not defined within the problem formulation. Therefore, the profit for selling energy to the grid is chosen lower than cost of buying energy, which encourages the optimizer to limit the absolute value of both variables.

Moreover, the chosen formulation $c_{Feed}(t) = 0.9 * c_{Buy}(t)$ has an additional reason: Since no charging or discharging efficiencies and therefore no losses are considered, the 10% lower selling prices ensures that trading energy is only performed when the predicted price difference is higher than 10%.

After the optimizer calculated the charging profile for the week, the SoC of the remaining NRMM and the BESS are used as the starting point for next week's optimization. This process is repeated for 51 weeks, and the electricity costs of each week are cumulated and added to the component costs for the total yearly costs $c_{Tot}$ described in equation (2).

This approach includes two major simplifications. The first one is that the optimizer has full knowledge about the upcoming week. This includes full knowledge on the dynamic day-ahead prices and exact arrival and departure times of the NRMM and their SoC for the whole week. Secondly, no efficiencies for the energy conversion losses are considered for the simulation, assuming 100 % efficiency of each component.

## 2.2 Outer layer

As described in Figure 1, the optimization process is structured into two layers: an inner layer focused on operation strategy optimization and an outer layer leveraging BO for broader component exploration. The previously described inner-layer optimization, focusses on reducing the yearly electricity costs $\sum c_{Elec}(t)$. Following this optimization, the component costs of this run $c_{Comp}$ are calculated and fed back to the outer layer, as the total cost $c_{Tot}$ or objective according to equation (2). This layer then employs bayesian optimization to explore the multi-dimensional parameter space effectively.

The BO is an active learning algorithm utilized to efficiently navigate various component topologies without the need for pre-training a constrained neural network. BO utilizes Gaussian Process regression to predict potential improvements in component parameters. This approach not only balances exploration and exploitation but also selects parameter sets based on expected improvement metrics, guiding the search towards optimal configurations. The computational efficiency of BO is particularly noteworthy, as it minimizes the number of iterations required to reach an optimized solution.

The BO limit was set to 100 iterations, while the first six configurations are chosen randomly to initially train the model. In order to limit the exploitation space to achievable configurations, the parameters were restricted. For each use-case, every EVSE was available, while the starting EVSE configuration was 11kW unidirectional charger. The allowed exploration range of the PV, BESS and their starting point are displayed in Table 2

Table 2: Possible exploration range and starting point of peak PV power, BESS capacity and max. grid power

| Components | Min | Start | Max |
|---|---|---|---|
| $P_{PV,Peak}$ | 0 kWp | 50 kWp | 200 kWp |
| $E_{BESS}$ | 0 kWh | 50 kWh | 100 kWh |
| $P_{Grid, max}$ | 20 kW | 150 kW | 300 kW |

In addition to the BO model learning the influence of the component parameters on the resulting objective, an additional error model is trained. The condition leading to an error in this simulation is the exceeding of the maximal grid power. Some possible combinations even enforce exceeding the power limit in order to charge the NRMMs to 100% (e.g. 20kWp peak grid power without a BESS).

To avoid a simulation error in these cases, the SoC target of the NRMM at departure time is implemented as a hard constraint within the simulation, while the exceeding of the designed peak power is penalized with high costs by the linear programming solver, effectively implementing a soft constraint. After the inner layer optimization, the compliance of the grid power with the predefined maximal grid power is checked and fed back to train the error model of the BO. The BO therefore not only tries to minimize the cost, but also to find feasible regions within the parameter space.

For more information on the utilization of BO, already applied in other applications, the reader is referred to [19, 20].

## 3. RESULTS

In this chapter, the progress of the BO through its iterations and the power flows of one exemplary week with the best setup found by the BO are shown.

The progress of the found minimum and the estimated minimum objective over the number of iterations is shown in Figure 4:

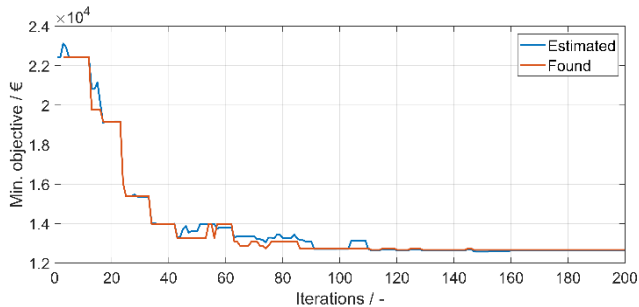

Figure 4: Bayesian optimization progress through the iterations

While the objective found during a simulation is the result of a performed simulation, the estimated minimum is the prediction of the BO which minimum objective can be achieved. Contrary to the found minimum, the estimated minimum is also considering the stability of a found configuration through the previously described error model. Therefore, the estimated objective can also be higher, if the found objective is considered too uncertain due to its proximity to a failed configuration. Towards the end of the training, the estimated and the found objective hardly differ. This is an indication for a well-trained BO model.

Even though 200 iterations were performed, it can also be noted that the found objective does not get reduced after the 111[th] iteration, which is also the final best configuration found by the BO. The competing Design of Experiment (DoE) approach would have needed 125 runs for simulating each possible charger configuration, considering the conservative approach to only simulate five options for each continuous value. The best configuration found by the BO and the corresponding costs for the 51 weeks simulated is shown in Table 3:

Table 3: Best configuration found by the bayesian optimization

| Component | Configuration | $c_{Comp}$ |
|---|---|---|
| PV | 133.9 kWp | 1575.3 € |
| BESS | 99.0 kWh | 1941.3 € |
| Grid, max | 69.4 kW | 7218.2 € |
| Use-Case 1 | EVSE 11 Bidi | 1569.2 € |
| Use-Case 2 | EVSE 11 Bidi | 523.1 € |
| Use-Case 3 | EVSE 11 Bidi | 523.1 € |

In order to investigate the optimized configuration by the BO, the power profile of two exemplary days from week 10 is shown in Figure 5:

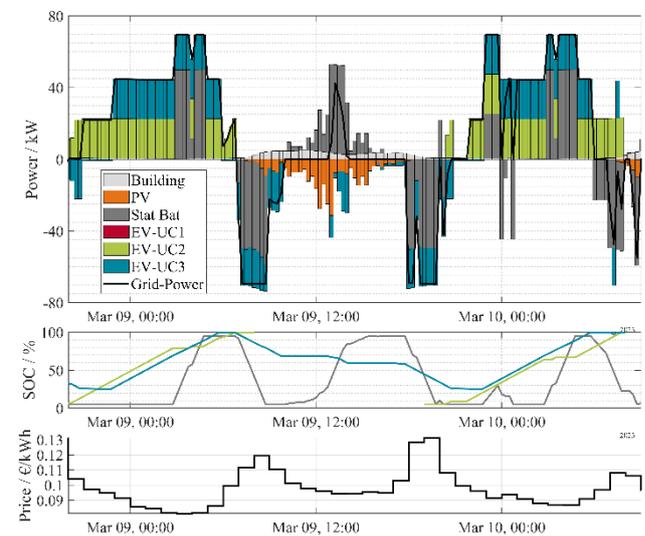

Figure 5: Exemplary power profile of Building, PV, BESS, NRMMs of each use case and grid power (top), averaged SoC of NRMMs from each use-case and BESS (middle) and electricity price (bottom)

The NRMM from UC-2 and UC-3 are depicted, with the NRMM from UC-2 just arriving at the rental building at 7pm. The shown price profile reflects the typical price profile throughout the day: Low prices around mid-day and mid-night and high prices in the morning and evening hours. So, the NRMM that only stay one night are already connected when typically, the lowest prices occur and are therefore just charged.

The NRMM from UC-3 on the other hand are only needed on the day after and act as a temporary storage similar to the BESS throughout the day. They are mostly charged during times with PV production and low prices and discharged at times with a high price. At around 8am on the first day, we can see that the BESS is discharged with its maximal power of 49.5kW and the 2 NRMM of UC-3 are discharged each with approximately 11kW, all while matching the grid limitation of 69.4kW due to the small consumption of the dealership building.

The desired SoC of 100% at departure time is achieved for all NRMMs throughout the simulation. In general, it can be seen that the BESS is prioritized to the NRMMs when discharging, since the NRMM need to be recharged later eventually. The same behavior of the NRMM from UC-3 is applicable to the NRMM from UC-1 and UC-2 when they are staying for more than one day on the weekend. Even though the possible time for discharging these NRMM is limited, the additional investment of 406€ for a bidirectional charging station is profitable.

The electricity cost for the shown two-day period in Figure 5 is 35.1€, while the cost for buying electricity is 72.6€ and the revenue for selling is 47.5€. The total charged energy for the NRMM is 515kWh, while the energy difference of the BESS can be neglected, as it is just charged 0.4% from the start to the end. This results in an average electricity price of 0.068€/kWh. Through the bidirectional charging stations, in combination with the BESS, the dealership even has a profit of 695€ for the electricity exchange with the grid throughout the year.

While these results seem promising, it needs to be mentioned that additionally fixed costs per kWh, due to taxes and renewable levies, need to be payed from dealership on top of the day-ahead energy price. As these fixed costs are not reimbursed when selling energy to the grid, a double taxation is applied to all EVs/NRMMs participating in energy trading through V2G, basically nullifying the economic potential. There are many initiatives that demand a reform of these regulation [21]. Without these double taxations, which is the premise of this research, the fixed costs would just have to be paid for every kWh that remains in the NRMM. The total offset by these fixed costs would be identical for every configuration due to the constant use-case and can therefore be neglected altogether when comparing different configurations.

## 4. CONCLUSION

This paper introduces a novel methodology that integrates Bayesian Optimization (BO) with an Energy Management System (EMS) to optimize electricity costs in a fleet rental service utilizing electric Non-Road Mobile Machinery (NRMM). The proposed approach demonstrates how V2G technology can be effectively applied within this context, showcasing its potential for enhancing energy flexibility and economic profits.

The study simulates a rental scenario involving ten electric tractors, incorporating bidirectional EVSE, a PV system, and a BESS. The inner layer of the optimization focuses on the operating strategy to minimize yearly electricity costs through dynamic charging profiles based on real-time grid prices. The outer layer employs BO to explore various energy component configurations efficiently, thereby optimizing both system performance and cost-effectiveness.

While the results seem promising, it is important to note that the use case was quite generic and the models employed were simplified. Further research is needed to refine the models and explore more complex scenarios. Additionally, assumptions regarding changing energy regulations were made, particularly the abolition of the double taxation for bidirectional EVs. It remains to be seen if these regulations will be enforced as they are crucial for the success of V2G.